*Article*

# Colour Perception in Immersive Virtual Reality: Emotional and Physiological Responses to Fifteen Munsell Hues

**Francesco Febbraio [1,2], Simona Collina [2], Christina Lepida [3] and Panagiotis Kourtesis [3,4,5,*]**

[1] Department of Classics, Sapienza University of Rome, 00185 Rome, Italy; francesco.febbraio@uniroma1.it
[2] Department of Education, Psychology and Communication, Suor Orsola Benincasa University of Naples, 80135 Naples, Italy; simona.collina@unisob.na.it
[3] Department of Psychology, The American College of Greece, 15123 Athens, Greece; clepida@acg.edu
[4] Department of Psychology, National and Kapodistrian University of Athens, 15784 Athens, Greece
[5] Department of Psychology, University of Edinburgh, Edinburgh EH8 9Y, UK
* Correspondence: pkourtesis@acg.edu

**Abstract**

Colour is a fundamental determinant of affective experience in immersive virtual reality (VR), yet the emotional and physiological impact of individual hues remains poorly characterised. This study investigated how fifteen calibrated Munsell hues influence subjective and autonomic responses when presented in immersive VR. Thirty-six adults (18–45 years) viewed each hue in a within-subject design while pupil diameter and skin conductance were recorded continuously, and self-reported emotions were assessed using the Self-Assessment Manikin across pleasure, arousal, and dominance. Repeated-measures ANOVAs revealed robust hue effects on all three self-report dimensions and on pupil dilation, with medium-to-large effect sizes. Reds and red–purple hues elicited the highest arousal and dominance, whereas blue–green hues were rated most pleasurable. Pupil dilation closely tracked arousal ratings, while skin conductance showed no reliable hue differentiation, likely due to the brief exposure times (30 s). Individual differences in cognitive style and personality modulated overall reactivity but did not alter the relative ranking of hues. Taken together, these findings provide the first systematic hue-by-hue mapping of affective and physiological responses in immersive VR. They demonstrate that calibrated colour shapes both experience and ocular physiology, while also offering practical guidance for educational, clinical, and interface design in virtual environments.

## 1. Introduction

Colour is one of the most salient dimensions of visual experience, shaping not only object recognition and depth perception but also emotional appraisal and behaviour. Psychological research consistently demonstrates systematic associations between hue and affective states. Warm hues such as red and yellow are typically linked with heightened arousal and dominance, whereas cooler hues such as blue and green tend to foster calmness, pleasure, and reduced stimulation [1–5]. These associations extend into cognitive and behavioural outcomes: red can increase attentional capture and avoidance motivation, while blue often facilitates creative thinking and approach behaviour [1,3].

Theoretical models of affect help situate these findings. Dimensional frameworks such as the Pleasure–Arousal–Dominance (PAD) model [1] describe emotional states across three continua, and the Self-Assessment Manikin [2] operationalises these dimensions through a pictorial, non-verbal rating tool. Such tools have become standard in colour–emotion research because they bypass linguistic biases while maintaining sensitivity to affective shifts. Importantly, colour also modulates autonomic physiology. Laboratory studies show that red environments can increase heart rate and skin conductance, whereas blue or green tones may reduce these indices [1]. Pupil diameter has emerged as an especially robust marker of arousal, reliably reflecting both luminance changes and emotional engagement [2]. Taken together, these strands of evidence indicate that colour exerts its influence through intertwined perceptual, affective, and physiological pathways, establishing it as a core variable in affective computing and human–computer interaction.

### *1.1. Colour Fidelity in Immersive VR*

The importance of colour is magnified in immersive virtual reality (VR). Unlike desktop displays, head-mounted displays (HMDs) occupy over 100° of the visual field, providing full-field stimulation that recruits extensive occipital networks [6], and magnifies behavioural impact [6]. Under these conditions, even minor chromatic deviations become highly noticeable and can disrupt presence, realism, and comfort [7] with colour shown to modulate affect and presence across immersive and desktop contexts [1,8,9]. Performance and perception are tightly coupled in immersive interaction, where attentional and engagement demands can modulate perceptual experience [10]. Empirical work confirms that subtle hue shifts can diminish plausibility, increase visual discomfort, and induce eyestrain [11]. Because every pixel contributes to the sense of "being there," colour errors that would be negligible on a flat screen can have significant psychological consequences in VR [12].

In applied settings, reliable colour reproduction is critical. Clinical VR platforms use colour palettes to regulate anxiety during exposure therapy, with calming hues acting as "safe zones" and warmer hues supporting the graded introduction of stressors [13]. In education, colour is used to highlight key information, direct attention, and enhance memory consolidation [8]. Safety training relies on colour-coded hazards and warnings, where fidelity is essential to support rapid decision-making [14]. If colours drift from their intended perceptual qualities—for example, if a cyan used to lower arousal desaturates toward grey–green—the intended psychological effect is undermined. Achieving colour fidelity in VR is technically complex. Human trichromatic vision relies on cones with peak sensitivities around 420, 534, and 564 nm [15], but the RGB primaries of HMDs only approximate these curves [16]. Classic accounts of human colour vision show that cone outputs are combined through opponent-process pathways [16–19], which introduce further constraints on perceptual coding. Additional distortions arise from the optical properties of the devices: the absence of ambient white points, reduced luminance ranges, chromatic aberrations between lenses, and inter-ocular hue shifts [20]. Even calibrated HMDs exhibit gamut deformation and spatial colour non-uniformities—especially in the periphery—highlighting the necessity of device-specific calibration [12]. Moreover, lens-induced transverse chromatic aberration introduces field-dependent hue shifts beyond ~5–6° eccentricity [20]. Recent audits of commercial HMDs have documented gamut compression and field-of-view–dependent hue shifts that exceed perceptual threshold [12,21]. In colour science, a ΔE00 value below 2 is widely considered the threshold for perceptual indistinguishability [22,23]. Given gamut deformation and field-dependent hue shifts reported in HMDs, colour management must target perceptual tolerances (e.g., ΔE00 < 2) with device-specific LUTs and imaging colorimetry [22,24,25].

Without calibration to this level, VR-based experiments risk systematic error, compromising both scientific validity and applied outcomes [22–25]. Imaging colourimetry and spectral correction pipelines are therefore indispensable for VR research on colour perception.

*1.2. Existing Evidence on Colour in VR*

Despite the importance of colour fidelity and the established role of hue in emotion, empirical evidence linking colour and affect in immersive VR remains limited. Systematic reviews confirm [8] that fewer than ten immersive studies have addressed this link, typically testing only a handful of hues and relying primarily on self-report. More recent work [11] has shown that colour effects on physiological arousal have been documented only for a few colour–lightness combinations (e.g., red vs. blue), underscoring both the scarcity and specificity of such findings.

Most have tested only two to six hues, often focusing on primary colours such as red and blue, and have relied primarily on self-report measures. For example, real-world and VR exposures across several colours have been compared, showing changes in cognitive performance and self-reported arousal but without collecting physiological data [26]. Another study contrasted red and blue VR rooms and, for the first time, measured heart rate and skin conductance, finding higher autonomic arousal in red environments [11]. While these results are informative, they provide only limited resolution.

Related work has also shown that colour appearance interacts with spatial perception in VR, with systematic differences between real and virtual rooms [25]. Additionally, in immersive environments, colour cues significantly modulate presence and emotional engagement [8].

Beyond these examples, a handful of studies have explored colour-emotion associations in semi-immersive or mixed-reality contexts, but they share similar constraints: small colour sets, inconsistent calibration, and reliance on verbal self-report [4,9,17]. None has systematically mapped a wide set of hues within an HMD, nor paired subjective ratings with concurrent autonomic measures such as pupil diameter and galvanic skin response. The absence of such work leaves a gap in both theory and practice. Theoretically, we lack a comprehensive hue-by-hue "emotional fingerprint" of immersive colour perception. Practically, VR designers and clinicians have little evidence to guide colour choices beyond general intuitions drawn from desktop studies. At the neural level, colour processing is supported by a distributed cortical network spanning V1, V2 and V4, which links perceptual experience to underlying mechanisms of visual coding, as comprehensively reviewed in classic accounts of cone-opponent and cortical pathways [18]. This provides a foundation for mechanistic interpretations of colour–emotion associations under VR constraints [19].

This study addresses this gap by providing the first systematic hue-by-hue mapping of affective and physiological responses in immersive VR. Our approach extends prior research by testing fifteen calibrated Munsell colours, ensuring perceptual fidelity ($\Delta E < 2$), and combining self-report ratings with autonomic measures (pupil size and skin conductance). In this way, we integrate colour science with immersive VR methods to produce a more comprehensive account of colour-emotion associations. The paper is organised as follows: Section 2 describes the materials and methods; Section 3 presents the results; Section 4 discusses theoretical, comparative, mechanistic, and applied implications; and Section 5 concludes with the main findings and directions for future research.

*1.3. Aims and Hypotheses*

The present study was designed to map emotional and physiological responses to a broad set of calibrated hues in immersive VR. Using the Munsell system, fifteen distinct

colours—ten chromatic and five achromatic—were presented within a head-tracked VR environment rendered through a high-fidelity HMD. Participants provided affective ratings along the PAD dimensions using the SAM, consistent with established desktop findings where warm hues elevate arousal and cool hues promote pleasure [1,27,28], while pupil diameter and skin conductance were recorded continuously to index autonomic arousal. To examine whether stable traits modulated colour reactivity, measures of cognitive style and personality were also collected.

We formulated two hypotheses:

**H1.** *Predicted that self-reported PAD ratings would vary systematically across hues, replicating established desktop findings but with potentially stronger effect sizes under immersive conditions.*

**H2.** *Predicted that physiological indices, particularly pupil dilation, would covary with hue in ways that mirror subjective arousal, while skin conductance might show weaker effects due to the brevity of exposures.*

By addressing these aims, this study establishes an empirical baseline for colour-aware VR design and provides a methodological bridge between calibrated perception, multidimensional emotion, and physiological responses, complementing cross-cultural evidence for universal but contextually shaped colour–emotion associations [9].

## 2. Materials and Methods

### 2.1. Participants

Thirty-six adults (18 women, 18 men; M = 26.0 years, SD = 5.5, range = 18–45) took part in the study. Participants were recruited from the American College of Greece (ACG) participants pool (students, staff, external volunteers, and recent alumni) via emails, social media, and noticeboard announcements. Testing took place from 19 March 2025 until 30 May 2025. All reported normal or corrected-to-normal vision and no colour blindness. Exclusion criteria included any neurological or psychiatric disorders, as well as the use of medications or medical conditions that could influence physiological or emotional responses. Eligibility was assessed through self-report during the recruitment process. Written informed consent was obtained digitally, and the study was approved by the Institutional Review Board of the American College of Greece (protocol #202503487, 18 March 2025). All 36 consented participants completed the study; there were no exclusions and no dropouts.

### 2.2. Hardware and Software

The experiment was conducted using a Varjo Aero VR head-mounted display (Varjo Technologies Oy, Helsinki, Finland; 2880 × 2720 pixels per eye, 115° field of view, 90 Hz refresh rate) paired with an HTC Vive Controller 2.0 (HTC Corp., Taipei, Taiwan) for participant input. Eye movements were recorded with the headset's integrated binocular infrared eye tracker, operating at 200 Hz with automatic five-point calibration. These apparatus choices are consistent with recommended standards for immersive VR neuropsychological testing, emphasising precision in tracking, reproducibility, and device selection [29]. Galvanic skin response (GSR) was collected using a Biosignals Plux 8-Channel Hub (PLUX Biosignals, Lisbon, Portugal) at 128 Hz, with electrodes placed on the volar surfaces of the index and middle fingertips. Experimental software was developed in Unity 2022.3-LTS (Unity Technologies, San Francisco, CA, USA; Windows 11 Pro, 64-bit) and run via Varjo Base 4.1 (Varjo Technologies Oy, Helsinki, Finland) with SteamVR 2.4 (Valve Corp., Bellevue, WA, USA).

All HMD firmware and software were updated to the latest stable releases as of April 2025. The virtual environment was authored in Unity 2022.3 LTS, launched via Varjo Base 4.1 with SteamVR 2.4. Scenes used unlit/self-emissive shaders (no scene lights) to avoid shading artefacts and keep apparent colour tied to shader inputs. The virtual environment consisted of a plain cubic room with diffuse-only wall shaders. Colour patches were rendered on the walls and calibrated once at the beginning of the study against printed Munsell references, achieving a colour accuracy of $\Delta E < 2$ [22,24]. This $\Delta E < 2$ threshold is consistent with colour science conventions [23] and with VR research linking perceptual fidelity in colour and scene realism [22,24].

### *2.3. Design and Stimuli*

The study employed a within-subjects design in which each participant was exposed to all experimental conditions. A total of fifteen colour stimuli were presented: ten chromatic hues, three grey levels, black, and white. Each stimulus was shown once per participant in random order. Stimuli were defined in the HSV colour space. To isolate hue effects, all chromatic stimuli were standardised at saturation of 75% and a value (brightness) of 60%, keeping saturation/brightness constant across hues. This differs from "pure" monitor primaries (e.g., HSV 120°, 100%, 50% for green) by design, to prevent saturation/lightness from confounding hue effects. This uniformity of saturation and brightness controlled for potential confounds of saturation and luminance on affective and physiological responses [4,11], in line with parametric models showing that lightness and saturation systematically modulate colour emotions and preference [27,28]. The achromatic stimuli (black, dark grey, medium grey, light grey, white) were defined with saturation set to 0% and value adjusted to produce the desired lightness levels.

During presentation, the entire virtual room (all six surfaces: four walls, floor, and ceiling) used unlit, self-emissive shaders with no scene lights, avoiding shading/shadow artefacts and isolating chromatic properties at the panel level. This created the perceptual impression of being fully surrounded by the hue, with no other visual objects or textures present. The descriptive labels, Munsell notations, and HSV coordinates of all 15 stimuli are presented in Table 1.

**Table 1.** Descriptive labels, Munsell notations, and HSV coordinates for the fifteen colour stimuli used in the experiment.

| Label | Munsell Notation | HSV Coordinates |
|---|---|---|
| Black | N 0/0 | 0°, 0%, 0% |
| Dark Grey | N 3/0 | 0°, 0%, 30% |
| Medium Grey | N 5/0 | 0°, 0%, 50% |
| Light Grey | N 7/0 | 0°, 0%, 70% |
| Red | 5R 5/6 | 0°, 75%, 60% |
| Yellow | 5Y 5/6 | 60°, 75%, 60% |
| Green | 5G 5/6 | 120°, 75%, 60% |
| Blue | 5B 5/6 | 240°, 75%, 60% |
| Purple | 5P 5/6 | 300°, 75%, 60% |
| Yellow–Red | 5YR 5/6 | 30°, 75%, 60% |
| Green–Yellow | 5GY 5/6 | 90°, 75%, 60% |
| Blue–Green | 5BG 5/6 | 180°, 75%, 60% |
| Purple–Blue | 5PB 5/6 | 270°, 75%, 60% |
| Red–Purple | 5RP 5/6 | 330°, 75%, 60% |
| White | N 10/0 | 0°, 0%, 100% |

*Note.* HSV values denote the input parameters to the rendering pipeline. For chromatic hues, S = 75% and V = 60% were fixed to keep saturation/brightness constant across hues (only hue varies).

For achromatics, S = 0% and V encodes the intended lightness step (N 3/0, N 5/0, N 7/0; black = N 0/0; white = N 10/0). This design intentionally departs from canonical "pure primaries" so that hue is the only manipulated dimension.

*2.4. Procedure*

After providing consent and completing the baseline questionnaires (Cognitive Style Index, BFI-2-S), participants were seated in the experimental room and fitted with the Varjo Aero XR headset and galvanic skin response electrodes, which were placed on the volar surfaces of the index and middle fingers of the non-dominant hand (see Figure 1). The headset was adjusted for comfort, and a five-point calibration routine in Varjo Base ensured eye-tracking accuracy within one degree of visual angle. Participants also wore closed-back headphones to minimise external noise and held the HTC Vive Controller 2.0 in their dominant hand, which was used only to provide ratings. They then completed a brief practice session (~3 min) in a neutral B&W VR room, which familiarised them with the headset, controller, and rating procedure. This grey environment also served as the physiological baseline.

During the experiment, each trial began with a 30 s white adaptation screen, followed by a 30 s presentation of one of the 15 colour stimuli. The entire cubic room (walls, floor, ceiling) was rendered in the target hue, producing the perceptual impression of being fully surrounded by colour. While the hue remained visible, participants rated their current state of Pleasure, Arousal, and Dominance on a –4 to +4 Self-Assessment Manikin scale using the controller. Ratings were self-paced, consistent with established colour–emotion protocols [1,2]. Pupil size and skin conductance were recorded continuously throughout each trial. The VR environment contained no other visual elements to minimise distraction. A full session, including the baseline and all 15 stimuli, lasted approximately 30 ± 2 min, and no breaks were requested. Figure 1 illustrates the experimental setup: participants were seated in front of a desk, wearing the Varjo Aero XR headset and holding the HTC Vive Controller 2.0, while electrodermal activity was recorded using fingertip electrodes connected to the biosignals hub.

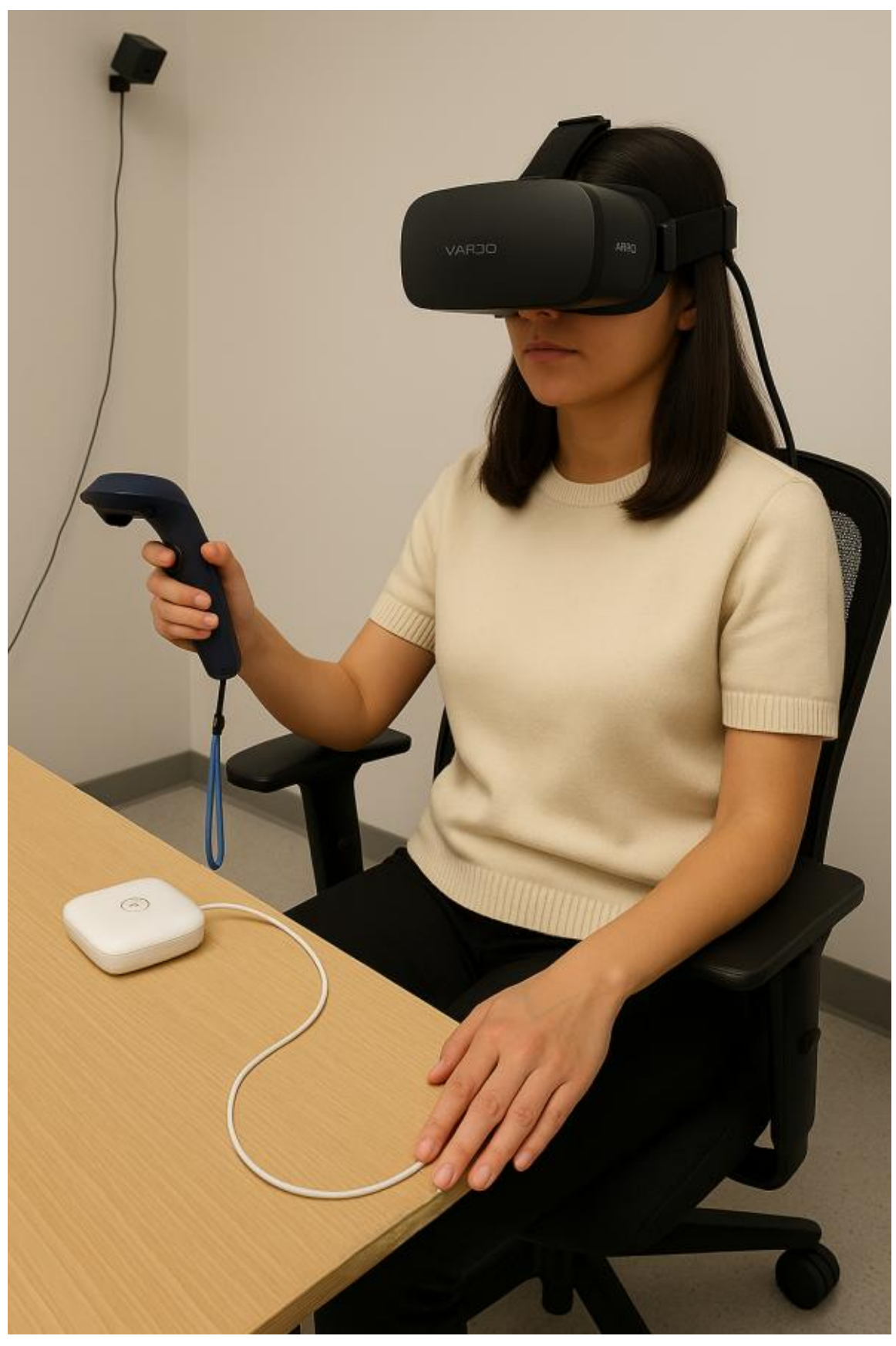

**Figure 1.** A participant during the experimental task, wearing the Varjo Aero XR head-mounted display (HMD) and holding the HTC Vive Controller 2.0. Electrodermal activity was recorded via fingertip electrodes connected to the Biosignals hub on the desk.

### *2.5. Measures*

We collected subjective emotional ratings, ocular and electrodermal responses, and trait-level individual differences. A full overview of variables, instruments, and timing is provided in Table 2.

**Table 2.** Descriptive overview of all measures, instruments, and timing used in the experiment.

| Domain | Variable | Instrument/Derivation | Timing |
| --- | --- | --- | --- |
| Subjective Emotion | Pleasure<br>Arousal<br>Dominance | 9-point Self-Assessment Manikin Sliders in VR Study Description | After Each Colour |
| Ocular Physiology | Pupil Diameter (mm) | Mean Value During Final 20 s of Colour Exposure | During Exposure to Colour |
| Electrodermal Activity | Skin Conductance (μS) | Mean and Peak Amplitude During Colour Exposure Minus Preceding White | During Exposure to Colour |
| Individual differences | Cognitive Style Index Total; Big-Five Inventory-2 Short Traits | Paper Forms pre-VR | Pre-Session |

#### 2.5.1. Subjective Emotion

Emotional responses were assessed using the Self-Assessment Manikin (SAM) [2] in combination with the Pleasure–Arousal–Dominance (PAD) framework [1]. The SAM is a non-verbal, pictorial instrument that presents stylised figures depicting varying levels of each affective dimension, allowing participants to indicate their current state quickly and intuitively. For each colour exposure, participants rated their levels of Pleasure (from unpleasant to pleasant), Arousal (from calm to excited), and Dominance (from submissive to in control) on a 9-point bipolar scale. Arousal ratings obtained with this format have been consistently shown to covary with autonomic responses such as pupil dilation and skin conductance [30]. This format has been widely validated in affective computing and colour–emotion studies. The SAM and PAD were adapted in a 3D version to be presented in VR, where responses were provided by touching the desired response (see Figure 2).

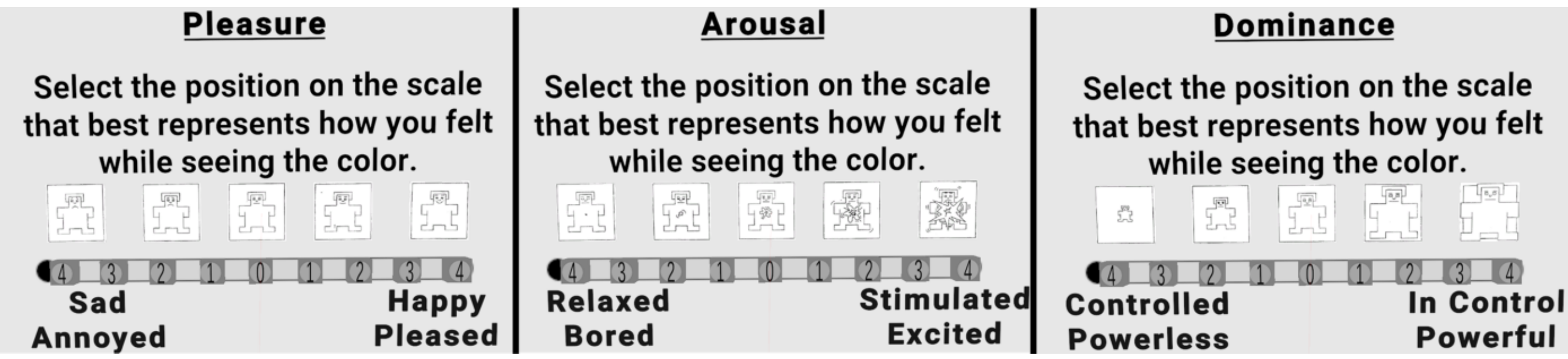

**Figure 2.** Illustration of the Self-Assessment Manikin (SAM) used for affective ratings of Pleasure, Arousal, and Dominance. Participants selected a position on a 9-point bipolar scale (–4 to +4) that best represented their emotional state while viewing each colour stimulus. This non-verbal format combines iconic pictorial figures with dimensional labels, reducing reliance on verbal interpretation.

#### 2.5.2. Pupil Dilation

Pupil diameter was recorded continuously through the Varjo Aero XR headset's integrated eye tracker. For analysis, the mean pupil size (in millimetres) was calculated over the final 20 s of each 30 s colour presentation. This approach minimised adaptation artefacts and provided a stable index of autonomic arousal, as pupil dilation reliably tracks both luminance and emotional state [31].

#### 2.5.3. Electrodermal Activity

Galvanic skin response (GSR) was recorded with a biosignalsplux 8-Channel Hub (PLUX Biosignals, Lisbon, Portugal) at 128 Hz. The hub provides up to eight synchronised analogue inputs at 16-bit resolution and supports per-channel sampling up to 2–3 kHz (higher when fewer channels are active) over Bluetooth Class II (~10 m line-of-sight range), powered by a 700 mAh Li-Po battery (≈10–12 h streaming). EDA sensors were connected to an analogue input; no separate reference lead is required for this sensor type. Channel configuration and live inspection followed the manufacturer's OpenSignals workflow (EDA/EDA.ARM presets), which provides signal previews and add-ons for phasic-tonic feature extraction. Electrodes (Ag/AgCl) were placed on the volar surfaces of the index and middle fingertips of the non-dominant hand; contact quality was checked before each block following the device's setup guidance. The hub integrates readily with Unity via the PLUX Unity API for device discovery and real-time acquisition; in this study, acquisition was run on the host PC and synchronised to Unity events. Mean and peak skin conductance (μS) during each colour trial were extracted, with values baseline-corrected against the preceding white adaptation screen. GSR is a sensitive measure of sympathetic activation and has been widely used to index arousal in psychophysiological studies [32].

#### 2.5.4. Individual Differences

To examine whether stable traits modulated colour reactivity, two questionnaires were administered prior to the VR session. The Cognitive Style Index (CSI) assesses individuals' preference for analytic versus intuitive modes of information processing, yielding both a total score and categorical style classification [33]. The Big Five Inventory-2 Short Form (BFI-2-S) provides brief but reliable measures of personality traits (Extraversion, Agreeableness, Conscientiousness, Negative Emotionality, and Open-Mindedness), as well as selected facets (e.g., Sociability, Trust, Energy Level) [34]. Both instruments are validated and widely used in cognitive and personality research.

#### 2.5.5. Cybersickness

Cybersickness was monitored using the Cybersickness in Virtual Reality Questionnaire (CSQ-VR), a brief six-item instrument validated for immersive VR (Cronbach's $\alpha \approx 0.86$) [35]. The questionnaire assesses three domains—nausea, vestibular disturbance, and oculomotor strain—each rated on a seven-point intensity scale. Subscale scores range from 2 to 14, and the total score ranges from 6 to 42, with higher values indicating more severe symptoms.

### *2.6. Data Pre-Processing*

Ocular data were screened for blinks (eye-openness < 50%), with affected frames removed; remaining pupil traces were smoothed with a five-sample median filter. Electrodermal signals were down-sampled to 10 Hz and log-transformed as log(μS + 1) to reduce skew. Unity event markers time-stamped stimulus onsets/offsets and were synchronised to biosignalsplux logs via host-PC timestamps; alignment was verified by spot-checking inter-event intervals. Streams were segmented into trials (30 s colour exposure following a 30 s white adaptation), and trial-level values were then aggregated within participants (e.g., means for continuous measures, peak amplitudes for skin conductance). Stimuli were rendered with unlit/self-emissive shaders (no scene lights), so no post-render image processing was applied; the specified HSV inputs directly controlled panel appearance. The resulting dataset comprised one observation per participant × hue for each dependent measure.

### *2.7. Statistical Analysis*

All analyses were conducted in R (version 4.4.0; R Foundation for Statistical Computing, Vienna, Austria) [36] using RStudio (version 2025.05.01; Posit PBC, Boston, MA, USA) [37]. Descriptive statistics and Bonferroni-adjusted correlation matrices were computed with psych [38]. Exploratory analyses included Pearson correlations between physiological measures and PAD ratings, as well as regression models in which CSI scores and BFI-2-S traits predicted mean Pleasure. Regression modelling was implemented with lme4 [39]. Confirmatory factorial tests used two complementary approaches. First, repeated-measures ANOVAs were fit with afex [40] (within-subject factor Hue, 15 levels; between-subject factor CSI style: Analytic vs. Intuitive), applying Greenhouse–Geisser corrections (ε) when sphericity was violated; effect sizes are reported as partial eta squared ($\eta^2_p$). Second—and as our primary inference—linear mixed-effects models (lme4/lmerTest) [39,41] were fit for each continuous outcome with fixed effects of Hue, CSI, and Hue × CSI, and by-participant random effects specified as (1 + Hue | participant_id); where the maximal structure was singular, a documented fallback (1 | participant_id) was used. Omnibus tests for Hue, CSI, and Hue × CSI were obtained via drop-one likelihood-ratio tests on nested models; fixed-effect variance explained is summarised with Nakagawa's marginal/conditional $R^2$ (performance). Estimated marginal means (EMMs) and pairwise contrasts were obtained with emmeans [42]; *p*-values were Holm-adjusted (Bonferroni for large contrast families as noted), with Satterthwaite degrees of freedom supplied by lmerTest. Model diagnostics comprised residual/fitted and QQ inspections and heteroscedasticity/influence checks (performance::check_model), simulated-residual tests (DHARMa), and collinearity indices (VIF/GVIF via performance::check_collinearity). "Baseline" trials (white adaptation) were excluded from inferential models; continuous variables were z-scored where appropriate, and when distributions deviated from normality, transformations via bestNormalize [43] were applied before modelling. As a robustness check, aligned-rank transform models were estimated with ARTool [44] using the same factorial structure, and post hoc contrasts were re-evaluated with emmeans. All tests were two-tailed with $\alpha = 0.05$.

## 3. Results

### *3.1. Descriptive Statistics*

Descriptive statistics for all measures are summarised in Table 3. The sample (N = 36) was gender-balanced (18 women and 18 men) and ranged in age from early adulthood (19 yrs) to midlife (45 yrs). Cybersickness was minimal. On the CSQ-VR, all participants reported scores of 1 ("absent") or 2 ("very mild") on each symptom, with the majority of responses being 1. On the CSI, they displayed a balanced distribution of cognitive styles (18 Analytic, 18 Intuitive), offering an opportunity to examine colour responses across different processing preferences. On average, self-reported emotions were close to the midpoint of the Self-Assessment-Manikin scales, with moderate pleasure, arousal, and dominance ratings. Physiologically, participants' mean pupil diameter was 3.64 mm, and their electro-dermal activity averaged 5.16 μS across trials. Notably, neither sex nor age altered the overall pattern of colour effects, although age was modestly associated with smaller pupil size and lower EDA. A slight positive skew in skin conductance data was addressed via log transformation, and non-parametric aligned-rank-transform tests confirmed that significant hue effects were robust against residual assumption violations.

For completeness, Table 4 reports the mean and standard deviation of each measure for every colour stimulus, including pupil diameter and skin conductance (GSR).

**Table 3.** Descriptive statistics for demographic, psychometric, and physiological measures.

| Variable | M | SD | Min | Max |
|---|---|---|---|---|
| Age (years) | 25.86 | 6.02 | 19.00 | 45.00 |
| Education (years) | 15.50 | 1.63 | 12.00 | 18.00 |
| CSI total score | 46.39 | 10.13 | 28.00 | 66.00 |
| Pleasure | 5.44 | 2.15 | 1.00 | 9.00 |
| Arousal | 5.10 | 2.20 | 1.00 | 9.00 |
| Dominance | 5.52 | 2.09 | 1.00 | 9.00 |
| Pupil size (mm) | 3.64 | 0.75 | 1.42 | 6.77 |
| Electrodermal activity (μS) | 5.16 | 3.23 | 0.05 | 16.40 |
| Extraversion | 20.44 | 4.06 | 13.00 | 29.00 |
| Agreeableness | 24.03 | 3.89 | 14.00 | 30.00 |
| Conscientiousness | 20.47 | 3.97 | 10.00 | 28.00 |
| Negative emotionality | 16.33 | 4.62 | 8.00 | 25.00 |
| Open-mindedness | 23.69 | 2.66 | 18.00 | 30.00 |
| Sociability | 6.94 | 1.92 | 3.00 | 10.00 |
| Assertiveness | 6.89 | 1.75 | 2.00 | 10.00 |
| Energy level | 6.61 | 1.71 | 3.00 | 10.00 |
| Compassion | 8.50 | 1.63 | 4.00 | 11.00 |
| Respectfulness | 5.56 | 1.24 | 3.00 | 9.00 |
| Trust | 7.22 | 1.62 | 2.00 | 10.00 |
| Organisation | 6.42 | 2.15 | 2.00 | 10.00 |
| Productiveness | 7.11 | 1.66 | 2.00 | 10.00 |
| Responsibility | 6.97 | 1.17 | 5.00 | 9.00 |
| Anxiety | 6.19 | 2.19 | 2.00 | 10.00 |
| Depression | 4.75 | 1.89 | 2.00 | 10.00 |
| Emotional volatility | 6.06 | 1.87 | 2.00 | 10.00 |
| Aesthetic sensitivity | 7.11 | 1.63 | 4.00 | 10.00 |
| Intellectual curiosity | 8.47 | 1.26 | 5.00 | 10.00 |
| Creative imagination | 8.11 | 1.54 | 5.00 | 10.00 |

All self-report traits are scored on their original scales. Physiological measures are averaged across all colour trials.

**Table 4.** Means (M) and standard deviations (SD) for Pleasure, Arousal, Dominance, pupil diameter (mm), and skin conductance response (μS) across the 15 colour conditions.

| Colour | Pleasure M (SD) | Arousal M (SD) | Dominance M (SD) | Pupil M (SD) | GSR M (SD) |
|---|---|---|---|---|---|
| Baseline | 4.72 (2.28) | 4.25 (2.25) | 5.44 (1.98) | 3.04 (0.39) | 5.78 (4.07) |
| Black | 4.53 (1.75) | 5.11 (1.98) | 5.19 (2.65) | 5.17 (0.61) | 5.24 (3.26) |
| Dark Grey | 4.50 (1.63) | 4.06 (1.41) | 5.14 (2.13) | 4.04 (0.82) | 5.01 (3.07) |
| Medium Grey | 4.14 (1.48) | 4.03 (1.54) | 4.89 (1.69) | 3.68 (0.45) | 5.20 (3.11) |
| Light Grey | 4.72 (1.78) | 3.89 (1.97) | 5.06 (1.84) | 3.16 (0.38) | 5.02 (3.43) |
| Red | 6.08 (1.93) | 7.25 (1.36) | 6.72 (1.91) | 4.06 (0.54) | 5.01 (3.16) |
| Yellow | 4.25 (2.33) | 4.56 (1.92) | 4.64 (2.26) | 3.43 (0.50) | 5.41 (3.81) |
| Green | 6.25 (1.93) | 5.53 (2.34) | 5.97 (1.84) | 3.51 (0.53) | 5.23 (3.44) |
| Blue | 6.75 (1.78 | 5.39 (2.48) | 6.28 (2.11) | 3.72 (0.50) | 5.41 (2.82) |
| Purple | 5.86 (1.84) | 6.28 (1.86) | 5.28 (1.95) | 3.54 (0.44) | 4.77 (2.97) |
| Yellow–Red | 4.44 (2.49) | 4.44 (1.87) | 4.28 (1.73) | 3.78 (0.64 | 5.17 (3.14) |
| Green–Yellow | 5.36 (2.27) | 4.50 (2.26) | 5.11 (2.31) | 3.52 (0.57) | 5.25 (3.07) |
| Blue–Green | 7.67 (1.35) | 5.61 (2.28) | 6.69 (1.69) | 3.18 (0.53) | 4.73 (3.04) |
| Purple–Blue | 6.11 (2.16) | 5.56 (2.44) | 5.89 (1.89) | 3.78 (0.57) | 5.12 (3.41) |
| Red–Purple | 6.00 (1.71) | 6.42 (1.95) | 6.08 (1.90) | 3.83 (0.45) | 5.08 (2.94) |
| White | 5.67 (1.79) | 4.67 (1.96) | 5.64 (1.94) | 2.72 (0.52) | 5.09 (3.16) |

Pleasure, Arousal, and Dominance were rated on 9-point SAM scales; pupil diameter measured in millimetres; skin conductance response (GSR) values were log-transformed (log[μS + 1]) prior to statistical analysis but are reported here in original μS units for interpretability.

### *3.2. Correlational Highlights*

Bonferroni-adjusted Pearson correlation matrices (Table 5) revealed several theoretically relevant associations. First, regarding cross-modal alignment, subjective Pleasure was strongly correlated with Dominance ($r = 0.62$, $p < 0.001$) and moderately with Arousal ($r = 0.42$, $p < 0.001$), confirming the internal coherence of the PAD model in immersive VR contexts. Second, in relation to personality variables, baseline-corrected pupil size was positively associated with Extraversion ($r = 0.18$, $p = 0.010$) and Energy Level ($r = 0.18$, $p < 0.001$), and negatively with Trust ($r = -0.20$, $p < 0.001$). Participants with higher scores on Openness (BFI-2 Open-Mindedness) also tended to report greater Pleasure for Blue–Green and purple hues (simple-slope $r$s ≈ 0.30–0.41, $p$s ≈ 0.02–0.04), although these hue-specific effects did not survive correction for multiple comparisons.

Finally, cognitive style showed a modest effect: the total CSI score was positively associated with Arousal ($r = 0.17$, $p = 0.010$), but not with any physiological measures. This finding is consistent with prior evidence that analytic–holistic tendencies are more likely to influence self-reported affect than reflexive bodily responses.

**Table 5.** Bonferroni-adjusted Pearson correlations among main study variables.

| **Predictor/Outcome** | **r** | ***p*** |
|---|---|---|
| Pleasure—Dominance | 0.62 | <0.001 |
| Pleasure—Arousal | 0.42 | <0.001 |
| Pupil size—Extraversion | 0.18 | 0.010 |
| Pupil size—Energy Level | 0.18 | <0.001 |
| Pupil size—Trust | −0.20 | <0.001 |
| Openness—Pleasure (Blue–Green) | ~0.30 | 0.02 |
| Openness—Pleasure (purple) | ~0.41 | 0.04 |
| Cognitive Style Index total—Arousal | 0.17 | 0.010 |
| Cognitive Style Index total—Physiology | n.s. | 1 |

### *3.3. Hue Effects: Mixed-Model ANOVA*

Repeated-measures ANOVAs with factors Hue (15 levels, within) and CSI style (between) were conducted to test the central hypotheses . Greenhouse–Geisser corrections ($\varepsilon$) were applied where sphericity was violated.

For Pleasure, there was a robust main effect of Hue, $F(7.64, 259.79) = 12.79$, $p < 0.001$, $\eta^2_p = 0.23$, indicating that colour hue explained nearly a quarter of the variance in affective ratings. A significant main effect of CSI was also observed, $F(1, 34) = 7.69$, $p = 0.009$, suggesting that participants with more holistic cognitive styles tended to report higher overall Pleasure. The Hue × CSI interaction was significant, $F(7.64, 259.79) = 2.63$, $p = 0.010$, $\eta^2_p \approx 0.07$, showing that the influence of hue on Pleasure differed according to individual differences in cognitive style.

For Arousal, Hue again exerted a strong effect, $F(7.65, 260.17) = 10.54$, $p < 0.001$, $\eta^2_p = 0.20$, accounting for one-fifth of the variance. CSI also predicted Arousal ratings, $F(1, 34) = 9.10$, $p = 0.005$, $\eta^2_p \approx 0.21$, with more analytic styles associated with lower Arousal. However, the Hue × CSI interaction did not reach significance.

For Dominance, a significant main effect of Hue was found, $F(7.77, 264.22) = 5.97$, $p < 0.001$, $\eta^2_p = 0.12$, demonstrating that hue systematically modulated participants' sense of control within the VR environment. No main effect of CSI was observed, but the Hue × CSI interaction was significant, $F(7.77, 264.22) = 2.91$, $p = 0.004$, $\eta^2_p \approx 0.08$, indicating that cognitive style moderated hue-driven differences in Dominance.

For pupil size, there was a very strong effect of Hue, $F(5.49, 186.58) = 95.31$, $p < 0.001$, $\eta^2_p = 0.49$, showing that nearly half of the variance in baseline-corrected pupil diameter was attributable to hue differences. Neither CSI nor the interaction term was significant. Finally, skin conductance did not show any reliable effects of Hue, CSI, or their interaction (all *p*s > 0.10).

As a robustness check against residual assumption violations, we also fitted aligned rank transform (ART) factorial models for each dependent variable. Table 6 reports the Type III omnibus tests (main effects and interactions); the ART results converged with the parametric ANOVAs and thus justify the planned pairwise contrasts reported below.

Post hoc analyses were conducted for each dependent variable to identify which hues differed significantly. Estimated marginal means were extracted from the mixed-model ANOVAs, and pairwise contrasts were tested using Holm-adjusted *p*-values. For each measure, hues were ranked according to participant-level means, and significant differences are detailed in Tables 7–9. As a summary, the brightest and most saturated hues (particularly yellow and cyan) tended to elicit the highest levels of Pleasure and Arousal, whereas darker hues (such as brown and dark grey) were consistently ranked lowest. For pupil size, short-wavelength hues (blue and purple) were associated with the largest dilations, whereas long-wavelength hues (red and orange) produced smaller responses, consistent with known spectral sensitivity patterns.

Figures 3–5 depict the full distribution of estimated marginal means (z-scaled) across all 15 hue stimuli and the baseline condition, offering a complete visualisation of hue effects. For transparency and reproducibility, extended tables with the full set of pairwise contrasts, exact *p*-values, and test statistics are provided in the Supplementary Material, allowing readers to inspect the complete statistical output beyond the summary rankings presented here.

**Table 6.** ART factorial tests (Type III) for each dependent variable.

| Dependent Variable | Effect | df | Sum of Squares | F | *p* |
|---|---|---|---|---|---|
| Pleasure | Colour | 15 | 3,756,838.89 | 13.117 | 0.000 |
| Pleasure | Cognitive Style | 1 | 659,757.29 | 7.444 | 0.010 |
| Pleasure | Interaction | 15 | 1,019,064.57 | 2.938 | 0.000 |
| Arousal | Colour | 15 | 3,331,229.89 | 11.480 | 0.000 |
| Arousal | Cognitive Style | 1 | 758,661.66 | 8.133 | 0.007 |
| Arousal | Interaction | 15 | 668,401.30 | 1.919 | 0.019 |
| Dominance | Colour | 15 | 2,050,373.67 | 6.652 | 0.000 |
| Dominance | Cognitive Style | 1 | 263,640.82 | 2.509 | 0.122 |
| Dominance | Interaction | 15 | 969,934.73 | 2.886 | 0.000 |
| Pupil size | Colour | 15 | 7,364,736.28 | 77.662 | 0.000 |
| Pupil size | Cognitive Style | 1 | 522,173.90 | 1.896 | 0.178 |
| Pupil size | Interaction | 15 | 289,065.11 | 1.632 | 0.061 |
| Micro siemens | Colour | 15 | 96,452.22 | 0.990 | 0.465 |
| Micro siemens | Cognitive Style | 1 | 267,598.17 | 0.741 | 0.395 |
| Micro siemens | Interaction | 15 | 151,049.99 | 1.566 | 0.079 |

Interaction = the interaction effect of colour and cognitive style.

#### 3.3.1. Pleasure

The Pleasure dimension showed significant variations across hues (Table 7, Figure 3). Blue–green emerged as the most pleasurable colour, with significantly higher ratings than 13 other hues, including the Baseline and all grey tones. Blue also scored highly, differing significantly from six other hues. At the lower end of the scale, Yellow was significantly lower than five other hues, while Medium Grey consistently occupied the lowest rank, differing significantly from eight other hues. These results indicate that brighter, more saturated hues—particularly those in the Blue–Green range—tend to evoke greater pleasure compared to less saturated or neutral tones.

**Table 7.** Highest and lowest ranked hues for Pleasure ratings, with descriptive statistics, number of significant contrasts, and *p*-value ranges.

| Rank | Colour | M (SD) | Significant Pairwise Differences with… | Total Colours | *p*-Value Range |
|---|---|---|---|---|---|
| Highest | Blue–Green | 7.67 (1.35) | Baseline, Dark Grey, Green, Green–Yellow, Light Grey, Medium Grey, Purple–Blue, Purple, Red–Purple, Red, White, Yellow–Red, Yellow | 13 | <0.001 to 0.033 |
| 2nd highest | Blue | 6.75 (1.78) | Baseline, Dark Grey, Light Grey, Medium Grey, Yellow–Red, Yellow | 6 | <0.001 to 0.004 |
| 2nd lowest | Yellow | 4.25 (2.33) | Blue–Green, Blue, Green, Red–Purple, Red | 5 | <0.001 to 0.003 |
| Lowest | Medium Grey | 4.14 (1.48) | Blue–Green, Blue, Green, Purple–Blue, Purple, Red–Purple, Red, White | 8 | <0.001 to 0.012 |

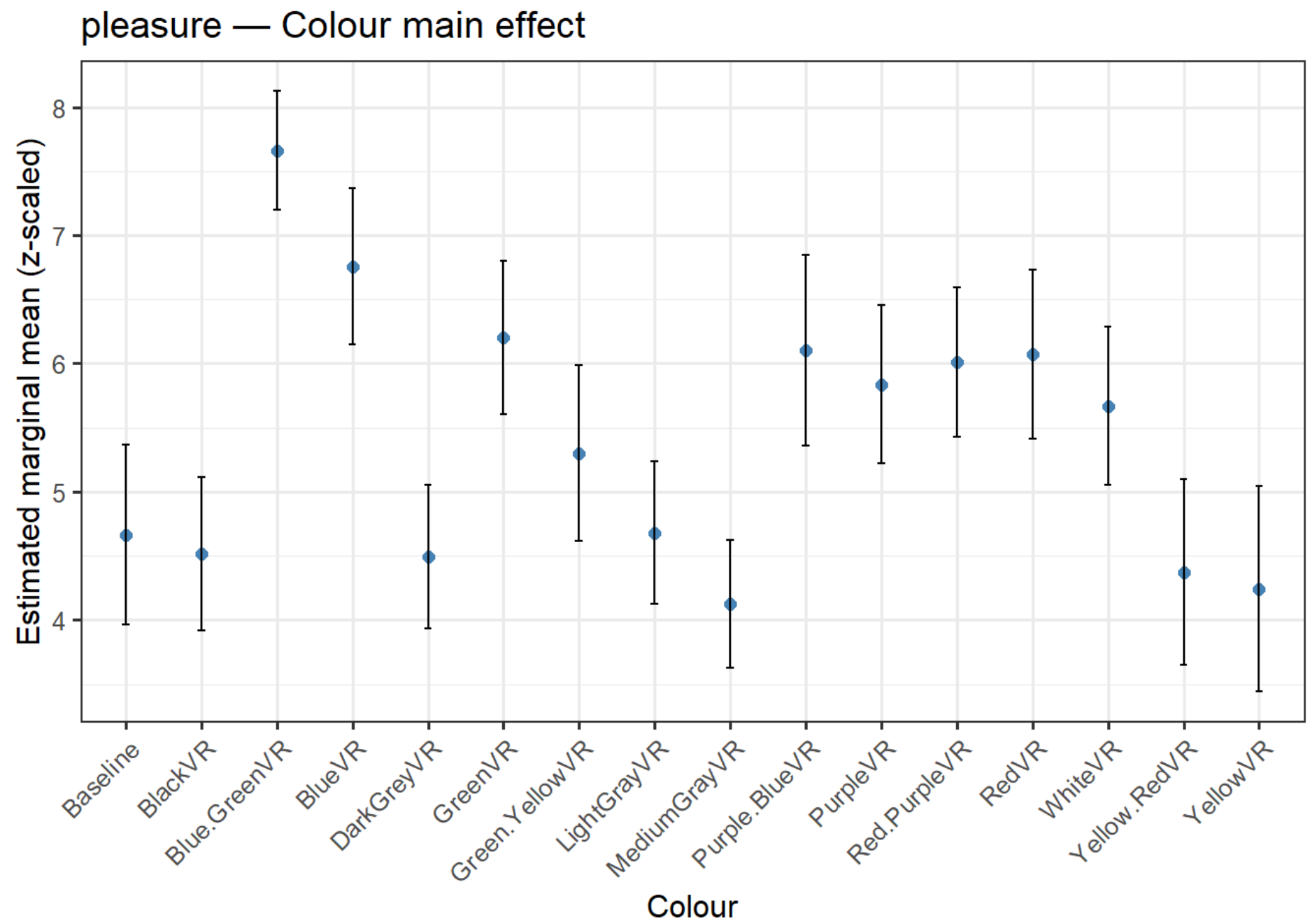

**Figure 3.** Estimated marginal means (z-scaled) for *Pleasure* across colour hues. Error bars indicate 95% confidence intervals. Colours are presented in HSV-defined order and include the Baseline condition. Highest pleasure ratings were observed for Blue–Green, followed by Blue; Medium Grey and Yellow elicited the lowest ratings.

### 3.3.2. Arousal

The Arousal dimension exhibited clear and robust differences between hues (Table 8, Figure 4), with warm and saturated tones occupying the upper end of the ranking. Red achieved the highest mean arousal score, showing significant pairwise differences with almost all other colours, including both neutral greys and cooler hues. Red–Purple and Purple also displayed high arousal values, significantly exceeding many of the lower-ranked colours. At the opposite end of the scale, Light Grey and Medium Grey elicited the lowest arousal levels, both being significantly lower than several high-arousal colours. This overall pattern suggests that hues in the red-to-purple range tend to evoke higher arousal, while desaturated greys are perceived as less stimulating.

**Table 8.** Highest and lowest ranked hues for Arousal ratings, with descriptive statistics, number of significant contrasts, and *p*-value ranges.

| Rank | Colour | M (SD) | Significant Pairwise Differences with… | Total Colours | *p*-Value Range |
|---|---|---|---|---|---|
| Highest | Red | 7.25 (1.36) | Baseline, Black, Blue–Green, Blue, Dark Grey, Green, Green–Yellow, Light Grey, Medium Grey, Purple–Blue, White, Yellow–Red, Yellow | 13 | <0.001 to 0.011 |
| 2nd highest | Red–Purple | 6.42 (1.95) | Baseline, Dark Grey, Light Grey, Medium Grey, Yellow–Red, Yellow | 6 | <0.001 to 0.033 |
| 2nd lowest | Medium Grey | 4.03 (1.54) | Purple, Red–Purple, Red | 3 | <0.001 |
| Lowest | Light Grey | 3.89 (1.78) | Purple–Blue, Purple, Red–Purple, Red | 4 | <0.001 to 0.048 |

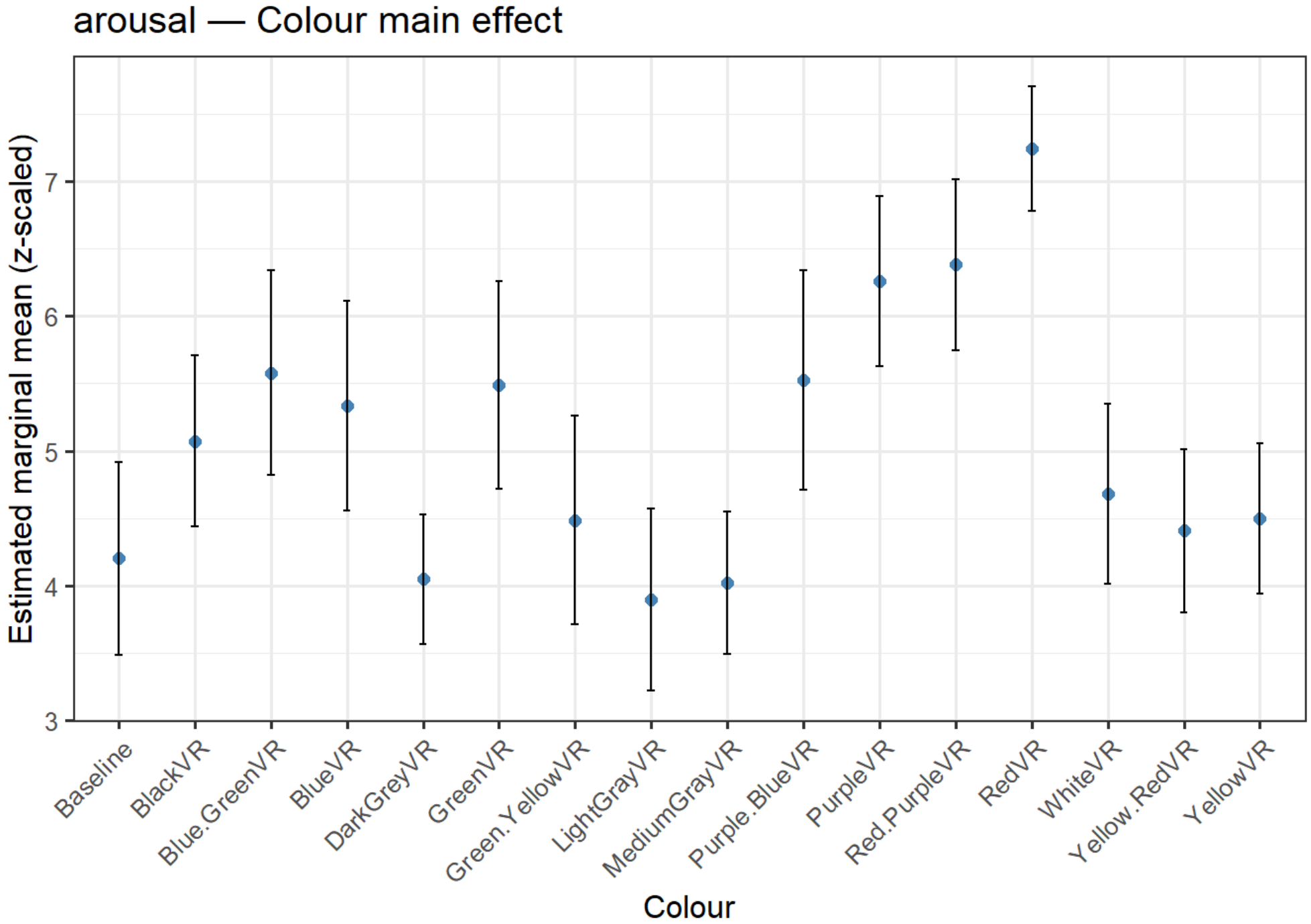

**Figure 4.** Estimated marginal means (z-scaled) for *Arousal* across colour hues. Error bars indicate 95% confidence intervals. Highest arousal ratings were observed for Red, followed by Red–Purple; Medium Grey and Light Grey elicited the lowest ratings.

#### 3.3.3. Dominance

The Dominance ratings positioned red and Blue–Green at the top of the ranking (Table 9, Figure 5), with mean scores that were not statistically distinguishable from each other ($p$ = 1.000). Both colours exhibited significantly higher dominance than several lower-ranked hues, particularly Yellow–Red, Yellow, and the grey tones. Blue followed in third position, showing significant differences from two of the lowest-ranking hues. At the lower end of the scale, Yellow and Yellow–Red consistently elicited the weakest dominance perceptions, being significantly lower than multiple high-ranking colours. These findings indicate that saturated primary hues, especially in the Red and Blue–Green range, are perceived as more dominant, whereas muted or composite hues, along with certain greys, tend to convey less dominance.

**Table 9.** Highest and lowest ranked hues for Dominance ratings, with descriptive statistics, number of significant contrasts, and $p$-value ranges.

| Rank | Colour | M (SD) | Significant Pairwise Differences with… | Total Colours | $p$-Value Range |
|---|---|---|---|---|---|
| Highest | Red | 6.72 (1.93) | Green–Yellow, Medium Grey, Yellow–Red, Yellow | 4 | 0.000 to 0.012 |
| Highest (tie) * | Blue–Green | 6.69 (1.35) | Green–Yellow, Light Grey, Medium Grey, Yellow–Red, Yellow | 5 | 0.000 to 0.023 |
| 2nd lowest | Yellow | 4.64 (2.33) | Blue–Green, Blue, Red | 3 | 0.005 to 0.039 |
| Lowest | Yellow–Red | 4.28 (2.49) | Blue–Green, Blue, Green, Purple–Blue, Red–Purple, Red | 6 | 0.000 to 0.010 |

* Red and Blue–Green occupy a joint first position in the dominance ranking. Although Red has a marginally higher mean score, the direct comparison between the two yielded a non-significant difference ($p$ = 1.000), indicating statistical equivalence in perceived dominance.

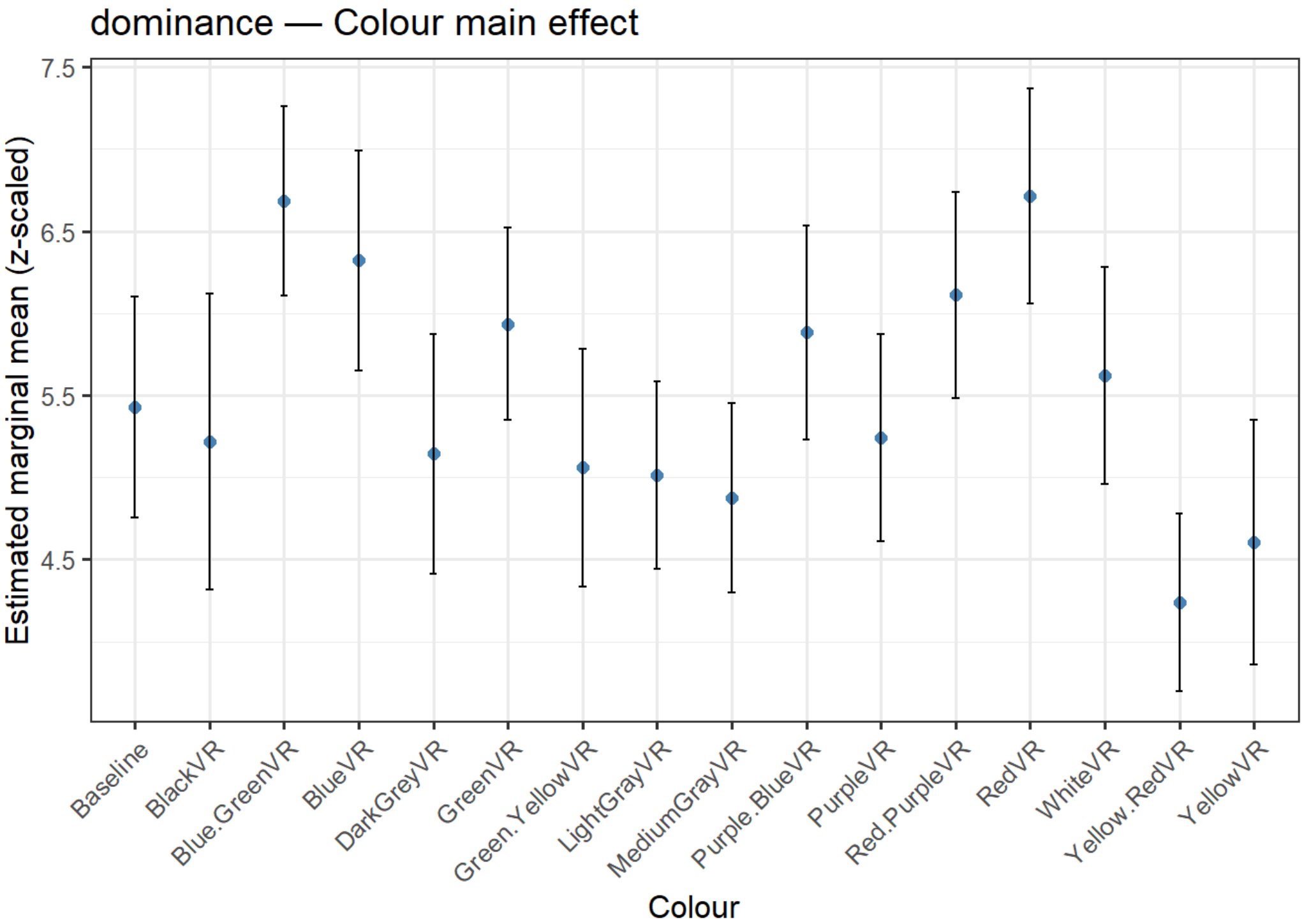

**Figure 5.** Estimated marginal means (z-scaled) for Dominance across colour hues. Error bars indicate 95% confidence intervals. The highest dominance ratings were observed for both Red and Blue–Green, whose mean difference was not statistically significant ($p$ = 1.000). Medium Grey and Yellow–Red elicited the lowest ratings.

### 3.3.4. Pupil Size

The analysis of pupil size revealed robust differences across hues (Table 10, Figure 6), reflecting both luminance-driven reflexes and, for certain colours, possible affective modulation. Black elicited the greatest pupil dilation, significantly exceeding almost all other hues, followed by Red, which also produced substantial dilation and matched the top-ranking Arousal colour. At the opposite end, White produced the smallest dilation, significantly lower than nearly all other colours, with the Baseline also ranking among the lowest values. This pattern suggests that darker, high-saturation hues—particularly Black and Red—trigger stronger pupillary responses, whereas lighter and neutral tones such as White and Light Grey evoke the weakest.

**Table 10.** Post hoc ranking of hues for Pupil Size, with mean (M), standard deviation (SD), and number of statistically significant pairwise differences relative to other colours.

| Rank | Colour | M (SD) | Significant Pairwise Differences with… | Total Colours | *p*-Value Range |
|---|---|---|---|---|---|
| Highest | Black | 5.17 (0.61) | Baseline, Blue–Green, Blue, Dark Grey, Green, Green–Yellow, Light Grey, Medium Grey, Purple–Blue, Purple, Red–Purple, Red, White, Yellow–Red, Yellow | 15 | <0.001 |
| 2nd highest | Red | 4.06 (0.54) | Baseline, Green, Green–Yellow, Light Grey, Medium Grey, Purple, Red–Purple, White, Yellow | 9 | <0.001 to 0.002 |
| 2nd lowest | Baseline | 3.04 (0.39) | Black, Red, Dark Grey, Red–Purple, Purple–Blue, Yellow–Red, Blue, Medium | 12 | <0.001 to 0.004 |
| | | | Grey, Purple, Green–Yellow, Green, Yellow | | |
| Lowest | White | 2.72 (0.52) | Black, Blue–Green, Blue, Dark Grey, Green, Green–Yellow, Light Grey, Medium Grey, Purple–Blue, Purple, Red–Purple, Red, Yellow–Red, Yellow | 14 | <0.001 |

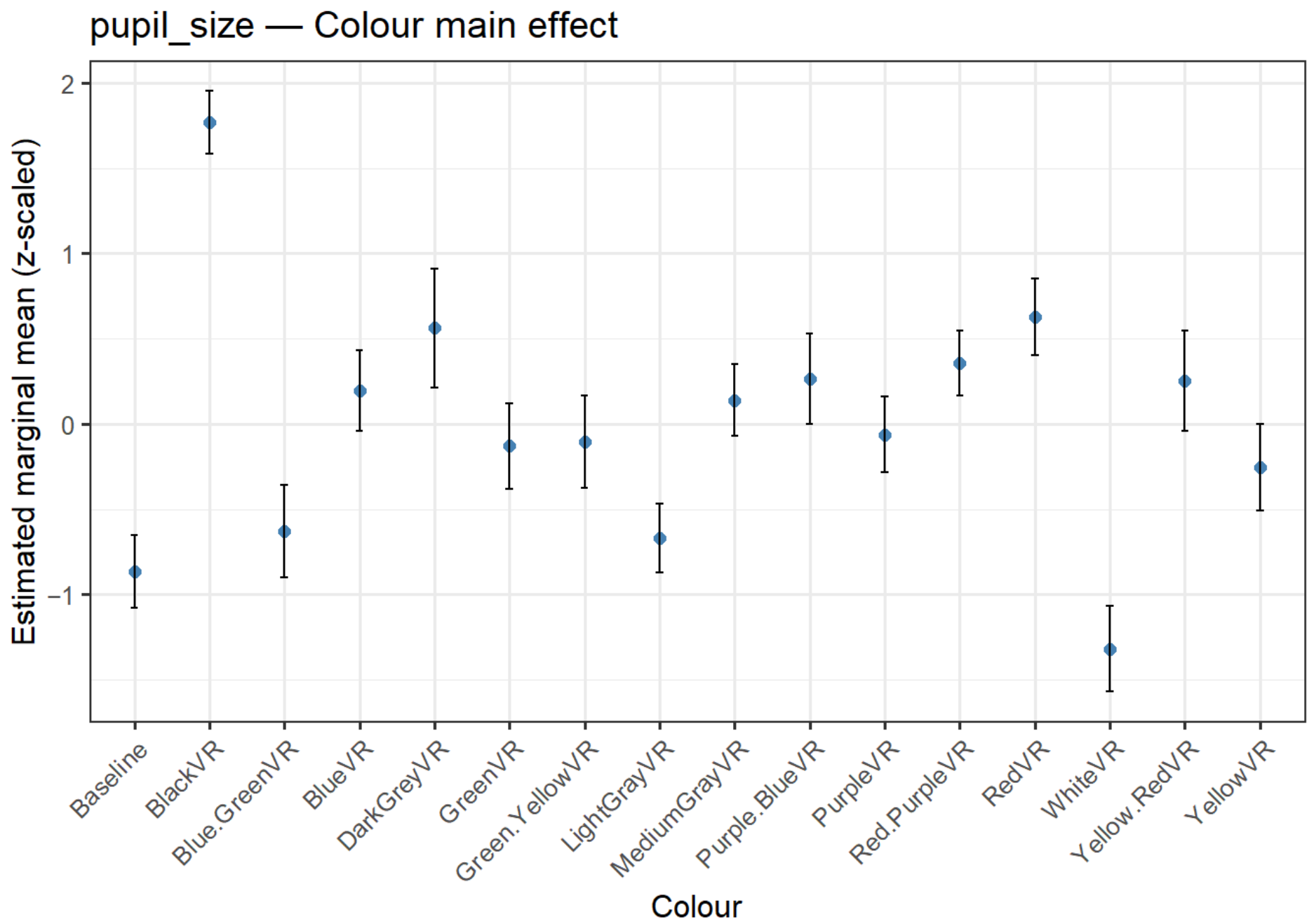

**Figure 6.** Estimated marginal means (z-scaled) for Pupil Size across colour hues. Error bars indicate 95% confidence intervals. The largest pupil dilation was observed for Black, followed by Red. White elicited the smallest dilation.

### *3.4. Hue Effects: Linear Mixed-Effects Models*

We fit linear mixed-effects models (LMMs) for each outcome with fixed effects of Hue (15 levels), CSI style (Analytic vs. Intuitive), and their interaction (Hue × CSI), plus by-participant random intercepts and random slopes for Hue (reduced to intercept-only where required). Likelihood-ratio tests (LRTs) compared reduced to full models (see Table 11). Model fit summaries (see Table 12) show modest-to-moderate fixed-effect signal for the self-report outcomes (Pleasure, Arousal, Dominance), a strong fixed-effect component for pupil size, and little fixed-effect signal for skin conductance; in all cases, conditional $R^2$ values were substantially higher, indicating considerable between-participant variability—especially for the physiological measures. All pairwise EMM comparisons use family-wise adjusted *p*-values (Holm/Bonferroni as specified). Full mixed-model outputs—LRT tables, fixed-effect estimates with 95% CIs, complete pairwise contrast matrices, and hue-ranking tables (overall and, where applicable, by-CSI)—are provided in Supplementary material.

**Table 11.** Omnibus likelihood-ratio tests from linear mixed-effects models

| Outcome | Effect | $\chi^2$ (df) | *p* |
|---|---|---|---|
| Pleasure | Hue | 196.38 (14) | <0.001 |
| | CSI | 52.08 (1) | <0.001 |
| | Hue × CSI | 47.07 (14) | <0.001 |
| Arousal | Hue | 165.50 (14) | <0.001 |

| | CSI | 42.75 (1) | <0.001 |
|---|---|---|---|
| | Hue × CSI | 35.95 (14) | 0.001 |
| Dominance | Hue | 128.68 (14) | <0.001 |
| | CSI | 58.02 (1) | <0.001 |
| | Hue × CSI | 56.35 (14) | <0.001 |
| Pupil size | Hue | 595.16 (14) | <0.001 |
| | CSI | ≈ 0.00 (1) | 1.000 |
| | Hue × CSI | ≈ 0.00 (14) | 1.000 |
| Skin conduct. (μS) | Hue | n.s. | ≥0.05 |
| | CSI | n.s. | ≥0.05 |
| | Hue × CSI | n.s. | ≥0.05 |

*Note.* Fixed effects: Hue (15 levels), CSI (Analytic vs. Intuitive), and Hue × CSI. Random effects: by-participant intercepts and (where supported) random slopes for Hue. $\chi^2$ tests compare the full model to reduced models (drop-1).

**Table 12.** Variance explained by the mixed models (Nakagawa $R^2$).

| Outcome | $R^2$m | $R^2$c |
|---|---|---|
| Pleasure | 0.277 | 0.380 |
| Arousal | 0.242 | 0.371 |
| Dominance | 0.174 | 0.326 |
| Pupil size | 0.473 | 0.814 |
| Skin conductance | 0.032 | 0.806 |

*Note.* $R^2$m = marginal (fixed effects only); $R^2$c = conditional (fixed + random effects).

#### 3.4.1. Pleasure

Hue exerted a robust effect, $\chi^2(14) = 196.38$, $p < 0.001$; CSI also contributed, $\chi^2(1) = 52.08$, $p < 0.001$; and Hue × CSI was significant, $\chi^2(14) = 47.07$, $p < 0.001$. Collapsing across CSI for a broad summary, the EMMs ranked Blue–Green and Blue at the top, with neutral greys near the bottom. Pairwise EMM contrasts mirrored this pattern: Blue–Green exceeded 13 of 14 alternative hues (including all greys), adjusted *p*s from <0.001 to 0.033; Blue was above six hues (mostly greys and warm tones), adjusted *p*s ≤ 0.004. At the low end, Medium Grey was significantly below eight higher-pleasure hues (e.g., Blue–Green, Blue, Green, Purple), adjusted *p*s ≤ 0.012, and Yellow was lower than five of the top hues (e.g., Blue–Green, Blue), adjusted *p*s ≤ 0.003. Together, these contrasts confirm a cool-saturated advantage (especially Blue–Green/Blue) and a grey penalty for perceived Pleasure (see Table 13).

**Table 13.** Top and lowest hues for pleasure: ranking based on Estimated-Marginal Means.

| Rank | Hue | EMM | 95% CI |
|---|---|---|---|
| 1 | Blue–Green | 7.64 | 7.00–8.28 |
| 2 | Blue | 6.74 | 6.10–7.38 |
| 3 | Green | 6.21 | 5.57–6.85 |
| 13 | Yellow–Red | 4.37 | 3.76–4.98 |
| 14 | Yellow | 4.24 | 3.63–4.85 |
| 15 | Medium Grey | 4.13 | 3.52–4.74 |

*Note.* EMM = Estimated-Marginal Mean; CI = Confidence Intervals.

#### 3.4.2. Arousal

The model again supported a strong effect of Hue, $\chi^2(14) = 165.50$, $p < 0.001$, and a CSI main effect, $\chi^2(1) = 42.75$, $p < 0.001$; the interaction was present, $\chi^2(14) = 35.95$, $p = 0.001$. Reflecting this, by-CSI rankings are informative. For Analytic participants, Red (highest) and Red–Purple reliably exceeded both greys and several cooler hues (e.g., Blue–Green, Blue), with adjusted *p*s ≤ 0.05; Light Grey and Medium Grey clustered at the bottom and were significantly below the high-arousal reds/purples (adjusted *p*s ≤ 0.05). For Intuitive participants, Red remained highest and exceeded the greys and multiple cool hues (adjusted *p*s ≤ 0.05). Overall, Arousal shows the expected warm-saturated boost (reds/purples) and a grey suppression (see Table 14).

**Table 14.** Top and lowest hues for arousal: ranking based on Estimated-Marginal Means.

| Rank | Hue | EMM | 95% CI |
|---|---|---|---|
| 1 | Red | 7.24 | 6.60–7.87 |
| 2 | Red–Purple | 6.38 | 5.74–7.02 |
| 3 | Purple | 6.26 | 5.62–6.89 |
| 13 | Dark Grey | 4.16 | 3.53–4.80 |
| 14 | Medium Grey | 4.07 | 3.44–4.71 |
| 15 | Light Grey | 3.77 | 3.14–4.41 |

*Note.* EMM = Estimated-Marginal Mean; CI = Confidence Intervals.

3.4.3. Dominance

Hue showed a reliable effect, $\chi^2(14) = 128.68$, $p < 0.001$; CSI was modest but present, $\chi^2(1) = 58.02$, $p < 0.001$; and Hue × CSI was significant, $\chi^2(14) = 56.35$, $p < 0.001$, indicating style-dependent modulation of perceived control. In the overall (CSI-collapsed) ranking, Red and Blue–Green occupied the top tier (see Table 15), each exceeding several lower-ranked hues—notably Yellow–Red, Yellow, and mid-greys (adjusted *p*s ≤ 0.023). Blue typically followed next, often exceeding one or more of the lowest group (e.g., Yellow, Yellow–Red; adjusted *p*s ≤ 0.05). At the low end, Yellow/Yellow–Red and mid-greys were reliably below the saturated primaries (See Table 15; Per-CSI simple-effects tables that reflect the significant interaction are included in the Supplementary material).

**Table 15.** Top and lowest hues for dominance: ranking based on Estimated-Marginal Means.

| Rank | Hue | EMM | 95% CI |
|---|---|---|---|
| 1 | Red | 6.13 | 5.50–6.77 |
| 2 | Blue–Green | 6.03 | 5.40–6.66 |
| 3 | Green | 5.78 | 5.15–6.41 |
| 13 | Light Grey | 4.33 | 3.70–4.96 |
| 14 | Yellow | 4.23 | 3.60–4.86 |
| 15 | Yellow–Red | 4.02 | 3.39–4.65 |

*Note.* EMM = Estimated-Marginal Mean; CI = Confidence Intervals.

3.4.4. Pupil Size and Skin Conductance

Hue had a very large effect, $\chi^2(14) = 595.16$, $p < 0.001$, whereas CSI and Hue × CSI were not reliable ($\chi^2(1) \approx 0$, $p = 1.000$; $\chi^2(14) \approx 0$, $p = 1.000$). The post hoc pattern is dominated by luminance/value: Black elicited the largest dilations (see Table 16), exceeding nearly every hue and all greys/White (adjusted *p*s < 0.001). Red also produced substantial dilation, surpassing many lighter or less-saturated colours (e.g., White, Light Grey, Yellow; adjusted *p*s ≤ 0.002). White showed the smallest dilation, significantly below almost all alternatives (adjusted *p*s < 0.001). This gradient—Black greater than cool hues greater than warm/light, with White lowest—matches the descriptive trend observed in ANOVA analyses and underscores the combined influence of luminance and hue (see Table 16).

For skin conductance (μS), fixed effects of Hue, CSI, and Hue × CSI were not reliable (all adjusted *p*s ≥ 0.05), while conditional $R^2$ remained high (≈0.81), suggesting pronounced individual variability without a consistent hue-linked fixed effect. Accordingly, we do not interpret pairwise hue contrasts for μS.

**Table 16.** Top and lowest hues for pupil size: ranking based on Estimated-Marginal Means.

| Rank | Hue | EMM | 95% CI |
|---|---|---|---|
| 1 | Black | 1.77 | 1.52–2.01 |
| 2 | Red | 0.63 | 0.38–0.87 |
| 3 | Dark Grey | 0.56 | 0.32–0.81 |
| 13 | Blue–Green | −0.63 | −0.87–0.39 |
| 14 | Light Grey | −0.67 | −0.92–0.43 |
| 15 | White | −1.32 | −1.56–1.08 |

*Note.* EMM = Estimated-Marginal Mean; CI = Confidence Intervals.

### *3.5. Summary of Findings*

The first hypothesis (H1) was confirmed: across both the ANOVAs and the mixed-effects models, hue reliably shaped subjective experience. The profiles were dimension-specific. Pleasure was highest for Blue–Green (followed by Blue) and lowest for Yellow and Medium Grey, consistent with a cool-saturated advantage and a penalty for desaturated greys. Arousal peaked for Red, with Red–Purple and Purple also elevated, whereas Light Grey and Medium Grey consistently fell at the bottom of the distribution, indicating that warm, saturated hues are perceived as more energising than greys or cooler tones. Dominance was strongest for Red and Blue–Green (their difference was not reliable), and weakest for Yellow and Yellow–Red, suggesting that saturated primaries convey a greater sense of control than composite or desaturated hues.

The second hypothesis (H2) received partial support. Pupil size showed a very large hue effect in the mixed models, with Black producing the greatest dilations, Red also relatively large, and White the smallest. This gradient accords with known luminance/value influences on the pupillary light reflex, with additional modulation by hue that broadly mirrors the arousal ordering. By contrast, skin conductance (EDA) did not exhibit reliable fixed effects of hue after baseline correction, implying that brief (30 s) passive exposures in an otherwise low-demand setting may be insufficient to consistently engage sudomotor activity even when self-reported arousal is high.

Regarding individual differences, the mixed-effects framework indicates that variability was captured primarily as scale differences rather than rank reversals. Participants with a more Analytic style tended to report slightly lower Pleasure and Arousal overall, and older participants showed smaller pupils, yet the relative ordering of hues was stable across sex, age, CSI style, and personality traits. In other words, people differed in how much they responded, not which colours they found most or least pleasant/arousing/dominant.

Taken together, these results demonstrate that in a head-tracked, colour-calibrated VR context, hue exerts robust, dimension-specific effects on subjective state and—via pupil diameter—on an objective physiological channel. The close coupling of Arousal with pupillary responses highlights pupil size as a practical, unobtrusive proxy for engagement in immersive media. At the same time, the absence of hue effects on EDA underscores that not all autonomic measures track chromatic stimulation equally under brief passive viewing, an important consideration for application domains (e.g., education, rehabilitation) that rely on physiological monitoring.

## 4. Discussion

To our knowledge, this is the first calibrated, hue-by-hue mapping of affective and physiological responses in immersive VR, and the conclusions are supported by both mixed-model ANOVAs and linear mixed-effects models (LMMs). Using fifteen calibrated Munsell hues, we found robust medium-to-large hue effects on subjective ratings of pleasure, arousal, and dominance. Pleasure was highest for Blue–Green and Blue, Arousal was strongest for Red and Red–Purple, and Dominance was greatest for Red and Blue–Green. Achromatic hues, particularly Medium Grey and Light Grey, were consistently rated lowest across affective dimensions. These effects were not only significant in omnibus ANOVAs but were replicated in LMMs that accounted for within-participant dependence and between-participant variability.

Physiologically, pupil dilation showed large hue effects that broadly paralleled arousal ratings: Black elicited the largest dilations, Red also produced substantial dilation, and White consistently produced the smallest responses. By contrast, electrodermal activity (EDA) did not differentiate among hues once the baseline was removed. Individual differences in cognitive style and personality modulated overall response magnitudes but did not alter the rank ordering of hues. Together, these findings demonstrate that colour influences both subjective states and pupillary responses in immersive VR, while GSR remains insensitive under short passive exposures.

The novelty of our study lies in three aspects: (i) the use of a high-end calibrated HMD to ensure colour fidelity, (ii) the inclusion of a broad and evenly spaced set of hues rather than a handful of "basic" colours, and (iii) the integration of self-report and physiological measures. These features allowed us to capture a fine-grained emotional "fingerprint" of colour in VR and to document stronger chromatic effects under immersive viewing than previously reported on desktop screens.

### *4.1. Theoretical Implications for Colour–Emotion Research*

Our findings provide new support for dimensional theories of affect, especially the PAD model [1], by showing that hue systematically modulates pleasure, arousal, and dominance in immersive contexts. The use of the Self-Assessment Manikin [2] enabled participants to make intuitive ratings without relying on verbal labels, reducing the risk of cultural or linguistic bias.

The pattern of results replicates and extends classic findings from flat-screen studies [1,4]. Warm hues increased arousal and dominance, while cool hues were associated with higher pleasure. The extension to immersive VR is theoretically important because full-field visual stimulation amplifies chromatic input, making colour a more salient contextual cue. The stronger effect sizes we observed suggest that colour–emotion links are not fixed constants but scale with the ecological validity of the viewing context.

Our physiological results also contribute to the psychophysiology of colour. Pupil dilation, often studied in relation to luminance and cognitive load [30,45], here emerged as a sensitive correlate of chromatic stimulation in VR, broadly tracking arousal ratings while also reflecting luminance/value. The moderate correlation between pupil change and self-reported arousal supports the idea that colour–emotion associations are embodied rather than purely symbolic [9]. In contrast, GSR showed no hue differences, underscoring that autonomic channels vary in sensitivity and timescale. Sweat gland activity may require longer or more emotionally intense exposures to differentiate among hues, with evidence that only saturated and bright colours elicit significant GSR changes during short (≈30 s) presentations [4].

### *4.2. Comparison with VR and Desktop Studies*

To date, few studies have directly examined colour–emotion links in VR. One investigation compared colour exposures in real and virtual environments, finding broadly similar effects on mood and performance but without measuring physiology [26]. Another contrasted red and blue VR rooms and reported higher heart rate and skin conductance in red environments [11]. However, most prior VR research has been limited to a small number of hues and relied primarily on self-report [8,11]. Our study extends this work by testing a much larger colour set, using calibrated stimuli, and integrating ocular physiology.

Unlike prior HMD studies that sampled only a handful of hues and relied primarily on self-report, we provide a hue-by-hue map across fifteen Munsell colours coupled with concurrent pupillometry—addressing the limitations highlighted in recent immersive work [11,26].

Compared with desktop research [1,27,28,46], our VR results showed amplified differences. For example, the Arousal gap between Red and Dark Grey was roughly double what has been reported on monitors. This amplification likely reflects the immersive properties of VR: a head-tracked, wide-field display fills the visual surround, eliminating external reference points and magnifying chromatic contrast [7]. Notably, physiological measures in VR also appear sensitive to environmental colour: darker rooms elicited increased heart rate and decreased heart rate variability, while red environments produced higher skin conductance than blue ones [11]; this highlights the nuanced role of hue and lightness in shaping affective responses.

Such amplification is not trivial; it suggests that theories of colour–emotion associations derived from desktop studies may underestimate the magnitude of effects in ecologically valid environments, while cross-cultural research highlights both universal regularities and contextual variations [9]. This aligns with evidence from immersive applications demonstrating that colour design significantly influences both the sense of presence and affective response in users [8].

By sampling fifteen evenly spaced hues, we were also able to detect fine-grained distinctions absent in previous work. For instance, Red–Purple elicited high arousal levels comparable to Red, while Blue–Green consistently produced the highest pleasure ratings. These subtleties illustrate the value of broad, calibrated hue sets for affective science. Our stimuli were uniform, full-field rooms—maximising chromatic immersion by engaging wide-field and peripheral vision with few competing cues. In real applications, colour is typically object- or UI-bound within cluttered scenes, where area/eccentricity, local and simultaneous contrast, material/BRDF, illumination/shading, and task demands all modulate appearance and affect. These factors generally dilute and localise the chromatic signal, so absolute effects will be smaller than our room-scale estimates, while the ordering of hues (e.g., Red > greys for arousal; Blue–Green > Yellow/Greys for pleasure) should largely hold. Future work should calibrate transfer from room- to object-scale colour rather than assume equivalence.

### *4.3. Mechanistic Interpretations*

Several mechanisms may account for the observed hue effects. One likely explanation involves cone-opponent pathways. Long-wavelength hues asymmetrically stimulate L- and M-cones, which increases perceived brightness and may activate the retino-hypothalamic pathway linked to sympathetic arousal. This account fits well with the heightened arousal ratings and pupil dilation observed for red and red–purple stimuli, consistent with psychophysiological evidence that arousal ratings covary with autonomic responses [30] and with cortical colour coding in extra-striate areas [19,47].

A second explanation concerns luminance-driven reflexes. The achromatic conditions clearly illustrate the role of lightness: black elicited the largest dilations and white the smallest, consistent with the pupil's primary role in regulating retinal illumination [30,31]. Because all chromatic hues in the present study were standardised to the same lightness level (V = 60%), differences in their pupil responses can be attributed mainly to hue rather than to luminance, although prior modelling has shown that lightness and saturation can strongly influence colour emotions and preference [27,28].

Another possibility relates to attentional salience. In a uniform VR scene, any deviation from a neutral baseline acts as an oddball stimulus, and longer-wavelength hues such as reds and yellows are especially likely to attract attention [48]. This heightened attentional capture may promote increased sympathetic drive and, consequently, greater pupil dilation. Such attentional-arousal pathways are consistent with evidence that performance and perception are tightly coupled in immersive contexts [10] and with findings that haptic feedback enhances vigilance and arousal during VR interaction [49].

Finally, device-specific factors may have played a role. The Varjo Aero's backlight exhibits a pronounced spectral peak near 456 nm and a secondary shoulder at 620 nm. If backlight spectra or sub-pixel layouts differentially weight long- vs. short-wavelength output, apparent brightness of certain hues could be biased. While our device was calibrated to target values, residual display characteristics may still modulate percepts, underscoring the need for multi-device replication. Such hardware artefacts highlight the importance of rigorous calibration in VR colour research [12,22,24].

Taken together, the results are best explained by an interaction of bottom-up sensory pathways, attentional salience, and device-level optics, each contributing to the observed hue-specific differences in affect and physiology. These findings are consistent with evidence from cortical colour networks in V1, V2, and V4, which provide the neural substrate for perceptual coding [19]. The parallel between arousal ratings and pupil dilation aligns with dimensional accounts of affect–autonomic coupling [30] and with immersive VR evidence showing that interaction conditions modulate perception and vigilance [10,45].

### *4.4. Practical Implications*

The present findings also carry several practical implications for the design of virtual environments. These recommendations are grounded in participant-level estimates from LMMs, increasing their robustness for deployment across heterogeneous user populations. In educational and training contexts, warm hues such as red and yellow can be employed sparingly to highlight feedback or mark critical elements, thereby boosting engagement, while cooler hues such as blue and blue–green can be used to sustain comfort and concentration during extended sessions [8].

In clinical VR, particularly for anxiety management or exposure therapy, blue–green palettes—rated here as least arousing—may serve as effective default "safe zones." Gradual transitions to warmer hues could then be used to support the graded introduction of therapeutic stressors, enabling clinicians to calibrate challenge and safety dynamically [13].

For interface and user experience (UX) design, our results highlight pupil dilation as a real-time, unobtrusive index of arousal [30,45]. Designers may leverage this signal to implement adaptive systems in which brightness or chroma levels are automatically adjusted to maintain engagement while reducing the risk of visual fatigue.

Finally, the results have implications for presence and comfort in immersive media. Careful colour calibration enhances perceptual realism, which in turn supports a stronger sense of presence and greater trust in VR systems [14,22]. By providing empirically grounded design principles, this study bridges basic colour science with applied VR practice, illustrating how calibrated palettes can be strategically used to shape emotional tone, manage cognitive load, and enhance user experience.

### *4.5. Limitations and Future Directions*

Several limitations of the present study should be acknowledged. First, the sample consisted of thirty-six young adults who experienced isolated colour rooms in a highly controlled laboratory setting. It remains unclear whether similar effects would be observed in older populations [7], in clinical groups, or in more content-rich and interactive VR environments that more closely resemble everyday applications.

Second, skin conductance proved insensitive under the present paradigm. The thirty-second passive exposures to static colour walls may have been too brief to reliably elicit electrodermal responses, which typically require stronger or longer-lasting stimulation. The absence of significant GSR differences should therefore not be interpreted as definitive evidence that hue has no effect on sympathetic activity; rather, it highlights the need for paradigms better suited to capturing slower autonomic dynamics.

Third, the findings were obtained using a single high-end headset. While the Varjo Aero provided excellent fidelity and rigorous colour calibration, outcomes may vary with other display technologies, particularly devices with lower pixel density or alternative illumination methods such as OLED. Generalisability across hardware platforms thus remains to be established. Establishing such generalizability requires adherence to software design principles and technological competency frameworks for immersive VR methods [29]. This consideration aligns with evidence that perceptual fidelity in colour strongly shapes scene realism in VR [22,24], underscoring the importance of calibration standards that can be generalised across display technologies.

Finally, another limitation pertains to ecological scope. Specifically, colouring the entire space likely sets an upper bound on chromatic influence, whereas most applications colour objects and UI elements. To bridge this, a concise next step is to (i) vary area × context (full-field, wall, object patch; empty vs. furnished scenes) under matched luminance/value, (ii) manipulate material/lighting (matte/gloss/metal; calibrated illumination/CCT), (iii) use eye tracking to compute gaze-weighted exposure, (iv) compare passive vs. goal-directed tasks, and (v) apply the same LMMs across room- and object-based paradigms. These tests will quantify attenuation/transfer from rooms to objects and yield design-ready coefficients for colour-aware VR.

Future studies should address these constraints in several ways. One priority is to implement dynamic colour gradients or temporal transitions to test whether changes in hue elicit stronger affective responses than static presentations. Another promising direction is the integration of VR with neuroimaging methods such as EEG or fMRI, which would allow mapping of cortical colour–emotion networks and clarify the neural substrates underlying these effects. Finally, applied research should explore adaptive colour-rendering systems capable of adjusting hues in real time to nudge users toward their preferred arousal state. Such systems could enhance engagement and emotional regulation while preserving comfort in long-duration VR sessions.

## 5. Conclusions

Immersive VR colours are more than decorative elements: they exert measurable influences on how users feel and how their bodies respond. In this study, reds were associated with the highest arousal and produced substantial pupil dilation, whereas blue–green tones elicited the greatest pleasure. Red and blue–green hues also conveyed a stronger sense of dominance. Black generated the largest pupil dilation overall, likely reflecting both heightened arousal and adaptation to low luminance, while white consistently produced the smallest dilation.

These findings suggest that VR creators can move beyond aesthetic considerations to employ evidence-based colour palettes that strategically shape emotional tone, manage cognitive load, and enhance the sense of presence. Warm, saturated hues may be used selectively to increase engagement, while cooler blue–green shades can help maintain comfort and reduce fatigue during prolonged exposure. By providing a systematic, hue-by-hue emotional and physiological map, this study establishes a foundation for colour-aware design principles in immersive virtual environments. By integrating calibrated hue maps with physiological metrics, our study offers a foundation for adaptive VR systems that adjust colour dynamically to optimise user comfort and engagement.

**Supplementary Materials:** The following supporting information can be downloaded at: https://www.mdpi.com/article/doi/s1. The Supplementary Materials contain the full analysis code (R Markdown and scripts), the processed analysis dataset with a variable dictionary, and a calibration dossier (colour pipeline, HMD settings, verification checklist). Beyond the PDF, an HTML version of the Supplementary Materials can be accessed here: https://rpubs.com/PanosCortese/1326834, accessed on 12 October 2025.

**Author Contributions:** Conceptualization, P.K. and F.F.; methodology, P.K. and F.F.; software, P.K.; validation, P.K., F.F., S.C. and C.L.; formal analysis, P.K.; investigation, F.F. and C.L.; resources, P.K. and S.C.; data curation, F.F. and C.L.; writing—original draft preparation, F.F. and P.K.; writing—review and editing, P.K., F.F., S.C. and C.L.; visualisation, P.K.; supervision, P.K.; project administration, P.K. All authors have read and agreed to the published version of the manuscript.

**Funding:** This research received no external funding.

**Institutional Review Board Statement:** The study was conducted in accordance with the Declaration of Helsinki and approved by the Institutional Review Board of the American College of Greece #202503487, 18 March 2025.

**Informed Consent Statement:** Informed consent was obtained from all subjects involved in the study.

**Data Availability Statement:** The original data presented in the study are openly available in Open Science Framework (OSF). The Unity build, project assets/executable, raw CSV exports, and ancillary files are archived at the https://osf.io/jw3dq/files/osfstorage , accessed on 8 October 2025.

**Conflicts of Interest:** The authors declare no conflicts of interest.

## Abbreviations

The following abbreviations are used in this manuscript:

| | |
|---|---|
| PAD | Pleasure–Arousal–Dominance |
| SAM | Self-Assessment Manikin |
| VR | Virtual Reality |
| CSI | Cognitive Style Index |
| BFQ-S | Big Five Inventory-2 Short Traits |
| GSR | Galvanic Skin Response |
| HMDs | Head-Mounted Displays |